\documentclass[12pt]{article}

\usepackage[utf8]{inputenc}
\usepackage{charter}
\usepackage{fullpage}
\usepackage{hyperref}
\usepackage{graphicx}
\usepackage{lipsum}
\usepackage[sort&compress,numbers]{natbib}
\usepackage{hyperref}
\hypersetup{hidelinks}
\usepackage[total={6.5in,9in},top=1in,headsep=0.1in,headheight=1in]{geometry}

\newcommand\snowmass{
\begin{center}
  \rule[-0.2in]{\hsize}{0.01in}\\
  \rule{\hsize}{0.01in}\\
  \vskip 0.1in
  Submitted to the Proceedings of the US Community Study\\ 
  on the Future of Particle Physics (Snowmass 2021)\\
  \rule{\hsize}{0.01in}\\
  \rule[+0.2in]{\hsize}{0.01in}\\[-2em]
\end{center}
}

\usepackage[firstpage=true]{background}
\backgroundsetup{contents={\parbox{6.5in}{\snowmass}}, scale=1,placement=top,opacity=1,color=black,position={3.25in,1.2in}}

\usepackage{fancyhdr}
\fancypagestyle{plain}{%
  \fancyhf{}%
  \fancyhead[C]{}
  \fancyfoot[C]{\thepage}
}

\fancypagestyle{empty}{%
  \fancyhf{}%
  \fancyhead[C]{{\it Snowmass2021 Cosmic Frontier White Paper: Rubin Observatory after LSST}}
  \fancyfoot[C]{\thepage}
}
\title{Snowmass2021 Cosmic Frontier White Paper: Rubin Observatory after LSST}
\date{}

\usepackage{authblk}

\author[1]{Bob Blum}
\author[2,3]{Seth W. Digel}
\author[4,5]{Alex Drlica-Wagner}
\author[6,7]{Salman Habib}
\author[7]{Katrin Heitmann}
\author[8]{Mustapha Ishak}
\author[9]{Saurabh W. Jha}
\author[2,3,10]{Steven M. Kahn}
\author[11]{Rachel Mandelbaum}
\author[2,3]{Phil Marshall}
\author[12]{Jeffrey A. Newman}
\author[2,3]{Aaron Roodman}
\author[13]{Christopher W. Stubbs}
 
\affil[1]{Vera C. Rubin Observatory/NSF’s NOIRLab 950 N Cherry Ave., Tucson, AZ, 85749, USA} 
\affil[2]{SLAC National Accelerator Laboratory, 2575 Sand Hill Rd, Menlo Park, CA, 94025, USA}
\affil[3]{Kavli Institute for Particle Astrophysics \& Cosmology, P.O. Box 2450, Stanford University, Stanford, California 94305, USA }
\affil[4]{Fermi National Accelerator Laboratory, P. O. Box 500, Batavia, IL 60510, USA}
\affil[5]{Department of Astronomy and Astrophysics, University of Chicago, Chicago, IL 60637, USA}
\affil[6]{Computational Science Division, Argonne National Laboratory, 9700 South Cass Avenue, Lemont, IL 60439, USA} 
\affil[7]{High Energy Physics Division, Argonne National Laboratory, 9700 South Cass Avenue, Lemont, IL 60439, USA}
\affil[8]{Department of Physics, The University of Texas at Dallas, Richardson, Texas 75080, USA}
\affil[9]{Department of Physics and Astronomy, Rutgers, the State University of New Jersey, 136 Frelinghuysen Road, Piscataway, NJ 08854, USA}
\affil[10]{Department of Physics, Stanford University, Stanford, CA 94305, USA} 
\affil[11]{McWilliams Center for Cosmology, Department of Physics, Carnegie Mellon University, Pittsburgh, PA 15213, USA}
\affil[12]{University of Pittsburgh and PITT PACC, 3941 O'Hara Street, Pittsburgh, PA 15260, USA}
\affil[13]{Department of Physics and Department of Astronony, Harvard University, 17 Oxford Street, Cambridge MA 02138, USA}

\begin{document}

\maketitle

\begin{abstract}
The Vera C. Rubin Observatory will begin the Legacy Survey of Space and Time (LSST) in 2024, spanning an area of 18,000 deg$^2$ in six bands, with more than 800  observations of each field over ten years. The unprecedented data set will enable great advances in the study of the formation and evolution of structure and exploration of physics of the dark universe. The observations will hold clues about the cause for the accelerated expansion of the universe and possibly the nature of dark matter. During the next decade, LSST will be able to confirm or dispute if tensions seen today in cosmological data are due to new physics. New and unexpected phenomena could confirm or disrupt our current understanding of the universe. Findings from LSST will guide the path forward post-LSST. The Rubin Observatory will still be a uniquely powerful facility even then, capable of revealing further insights into the physics of the dark universe. These could be obtained via innovative observing strategies, e.g., targeting new probes at shorter timescales than with LSST, or via modest instrumental changes, e.g., new filters, or through an entirely new instrument for the focal plane. This White Paper highlights some of the opportunities in each scenario from Rubin observations after LSST.    

\end{abstract}



\section{Introduction}\label{sec:intro}

The Vera C. Rubin Observatory\footnote{https://www.lsst.org/}, currently under construction on the Cerro Pachón ridge in north-central Chile, will embark on the 10-year Legacy Survey of Space and Time (LSST) in 2024. The survey will enable a rich set of scientific investigations, including studies of the nature of dark energy and dark matter, mapping of the solar system, exploring the transient optical sky, and surveying the Milky Way~\cite{2019ApJ...873..111I}. Rubin will enable groundbreaking studies of fundamental cosmology, with precision measurements of the growth of structure as parametrized in the $\Omega_M - \sigma_8$ plane, the nature of dark energy in the $w_0 - w_a$ plane, and cosmic expansion via the value of the Hubble parameter. LSST studies of the dark universe are of key importance to the Snowmass discussion. This portfolio includes investigations of dark matter in the cosmic context as presented in detail in Ref.~\cite{Rubin-DM-Snowmass}, including a number of probes of the particle nature of the dark matter~\cite{2019arXiv190201055D}. 

While the exact LSST survey strategy is still being optimized, the basic plan has been in place for many years. In short, LSST will consist of 1) a deep wide-field survey of roughly 18,000 deg$^2$ in the Galactic caps with repeated images taken on each part of the available sky every three or four nights, leading to at least 825 visits of every portion of the survey area in $ugrizy$ filter bands, 2) a set of four or five deep-drilling fields of 10 square degrees each with images taken approximately nightly, and 3) additional special surveys to cover the Galactic plane, Northern spur of the ecliptic plane, and the South celestial pole region, with fewer images collected.  Of course, Rubin Observatory is capable of other modes of operation, with either shorter or longer exposures than the currently planned pair of 15~second images per visit taken on each field, and either more or less repeated viewing, and a limited target of opportunity observing mode is being developed.  

The next decade and beyond promises to be very exciting for the field of cosmology. Stage IV dark energy surveys, including those undertaken by Dark Energy Survey Instrument (DESI), Rubin, the Nancy Grace Roman Space Telescope and the Euclid mission, will deliver unprecedented measurements of the growth of structure, and independently, the expansion rate of the universe. These measurements will shed light on the cause of the accelerated expansion of the universe and will be crucial for understanding inconsistencies already evident today. Currently, the so-called $H_0$ tension, now at the $5\sigma$ level~\cite{Hubble_tension,Garcia-Quintero:2019cgt}, points to a gap in our observational understanding of the universe or of the connection between the early and late-time epochs. Another developing tension in the  $\Omega_M-\sigma_8$ plane~\cite{2021A&A...646A.140H,Garcia-Quintero:2019cgt}, could present a similar puzzle if it endures in late-time measurements with higher precision soon to come from both DESI and Rubin.

Clearly the state of our understanding of cosmology in 2034 will inform what questions are of the greatest interest. We can look to the current state of cosmology for some guidance.  Should the present tensions persist, a next-generation project would very likely be focused on shedding additional light on the source of these discrepancies. Another future scientific driver is the level of precision in the $w_0 - w_a$ plane, along the lines of the Dark Energy Task Force~\cite{DETF} figure-of-merit. Stage IV projects DESI and Rubin are expected to improve the precision of constraints considerably over the current Stage III experiments. Certainly if a deviation is observed from $\Lambda$CDM in the equation of state, more precise or incisive measurements will be a high priority for understanding the nature of dark energy. Furthermore, $\Lambda$CDM remains a parametric description of our universe lacking deep underlying physical motivation. 
Over the next decade, experimental tests (including those performed by cosmic surveys) may provide insights into the fundamental constituents of this model (i.e., dark energy, dark matter, modifications to gravity, etc.), which will need to be interpreted in the context of the observed universe.

We discuss scientific opportunities for a range of possible additional surveys to further our understanding of dark energy, dark matter and cosmic expansion that could be realized with the considerable capabilities of Rubin Observatory $after$ completion of the LSST. Some of these have been discussed in Refs.~\cite{2019arXiv190504669S,2019BAAS...51g.273K,2019BAAS...51g.268J}. We consider continued operation of the observatory in three scenarios: 
\begin{itemize}
    \item Modifying only the observing strategy and possibly also observing entirely different regions of the sky than the LSST;
     
     \item Installing a new complement of filters for the camera; 
     \item Installing an entirely new focal plane instrument. This last scenario is obviously the most expensive, but could be the most compelling scientifically and should be investigated further. 

\end{itemize}

\section{Science Drivers}\label{sec:science}

Regarding cosmic acceleration and the dark energy associated with it, as well as dark matter, we are in an era of physics where much is known about the distribution of the components of the universe but very little is understood about their exact nature; moreover, tensions between different cosmological probes exist, and the field is still in the process of ensuring all relevant systematics are controlled at the necessary level.  LSST will allow the community to characterize with unprecedented precision the equation of state of dark energy, the astrophysical properties of dark matter, and also shed light on the current tensions. However, these results will only be the beginning rather the end of the quest to understand the dark universe and will open several science-driving questions for Rubin Observatory post-LSST:

\begin{itemize}
    \item If LSST and other surveys demonstrate with high precision that the equation of state of dark energy is different from a cosmological constant, i.e., $w_0 \ne -1$ and/or $w_a \ne 0$ then a profound opportunity arises to further explore the origin of this dynamical energy component. Targeted constraints could allow discrimination between various contenders with equations of state different from a cosmological constant based on new observational campaigns with Rubin.  

    \item In another scenario, confirming that the dark energy equation of state is essentially indistinguishable from $w_0 = -$1 and $w_a = 0$ will create the opportunity for post-LSST observations to explore dark energy beyond the equation of state approach. An example of such an approach is to pursue a sophisticated comparison between the expansion history and the growth rate of large-scale structure \cite{Ishak:2005zs,Ruiz:2014hma} to distinguish between vacuum energy or a modification to gravity that acts like dark energy.
    
    \item LSST will constrain to very high precision how light propagates in the universe using weak lensing and will reveal any inconsistency with our current theory of gravity. New observations after LSST would enable follow up on such results and provide unprecedented precision on constraining gravity on cosmological scales and the understanding of the nature of space and time.
    
    \item Within the next decade, direct, indirect, and/or collider experiments may detect evidence of particle dark matter. If so, astrophysical observations will be crucial to interpret those experimental results in the context of the observed dark matter distribution. Rubin observations after LSST will provide exceptional proper motion measurements to constrain the local density and velocity distribution of dark matter. On the other hand, if terrestrial experiments fail to detect dark matter, continued study of astrophysical probes with Rubin after LSST may be the most promising avenue for discovering and characterizing the fundamental properties of dark matter (e.g., through long-duration microlensing searches for primordial black holes).
    
    \item LSST has the potential to make the first statistically significant detection of dark matter halos that are too small to host galaxies. The existence of these halos is a definitive prediction of the $\Lambda$CDM model. Their detection would be a strong confirmation of $\Lambda$CDM, while a robust determination of the absence of such halos would demonstrate that $\Lambda$CDM is not an adequate description of the universe on very small scales. In either case, Rubin observations of stellar streams after LSST (e.g., with additional narrow band filters or improved proper motion estimates) would reduce contamination and strengthen conclusions about the existence of dark matter halos below the threshold of galaxy formation. 
    
    \item Last but not least, the apparent discrepancy of the amplitude of matter fluctuations, $\sigma_8$, between weak lensing surveys and cosmic microwave background (CMB) data, and the well-established $H_0$ tension, if they persist, are areas where the Rubin Observatory after LSST could play a major role in tracking down their origins regardless whether they are due to systematic effects in the data or result from new physics.
    
\end{itemize}

In short, independent of what is found in the next decade from LSST and other surveys, we will very likely not be done, in the sense of truly understanding the physics that underlies the accelerated expansion of the universe and the origin of dark matter. However, since the Rubin Observatory has not yet begun LSST survey operations, it is difficult to predict exactly what scientific questions will be considered critical to address post-LSST. The possibilities are numerous for how Rubin Observatory could enable science aligned with (potential future) DOE HEP priorities, and they will be refined by the results in the coming years. Exploring these options more deeply around year 5 of the 10-year survey, when the scientific landscape is clearer with initial results from LSST and DESI, would be valuable to identify how specifically this powerful facility can advance DOE HEP scientific interests in the post-LSST era.  

\section{Opportunities}

\subsection{Continuing Operations}
The extremely large etendue provided by the Simonyi Survey Telescope and the LSST Camera will remain a uniquely powerful combination after the completion of the LSST survey, and will continue to enable highly productive science, even without any modification to the hardware.  The depth and uniformity that will be achieved for LSST will be difficult to significantly improve upon with continued survey operations.  However, strong science cases for continued operation of Rubin relate to time domain studies that would rely on modified observing cadence, exposure time, or filter selections relative to the LSST survey for greatly enhanced efficiency and target-of-opportunity observations of rare phenomena. 

The prospects for precision measurement of $H_0$ via strong lensing cosmography are starting to be realized and efficient strategies for modeling strong lensing systems have been developed that make the method feasible for the relatively large numbers (hundreds) of strongly lensed quasars to be detected by LSST  \cite{2021ApJ...910...39P}.  The detection efficiency of the survey depends on fundamental aspects of the survey cadence, such as the  coverage of the extragalactic sky, season length, and filter selection.  The actual cadence for LSST will be a complicated optimization over all LSST science cases, and significant optimizations to increase the yield of ``golden'' lensed quasars are possible \cite{2021arXiv210405676L} and could be employed to further enlarge the sample of well-characterized systems post-LSST. The statistical precision of strong lensing cosmography will improve with the characterization of more systems, at least down to a precision of $\sim$0.2  km Mpc$^{-1}$ s$^{-1}$ \cite{2021ApJ...910...39P} -- potentially addressing the $H_0$ tensions more effectively.

Gravitational wave (GW) observatories are estimated to detect strongly gravitationally lensed neutron star-neutron star mergers at the rate of about 0.1 yr$^{-1}$ from the mid-2020s  \cite{2019arXiv190205140S}.  The kilonova optical counterpart will likely  be too faint to be detected in the LSST survey cadence.  Because detection of such a rare event would be a major advance, and because identification of the kilonova counterpart would be essential to establish its strongly lensed nature \cite{2019MNRAS.485.5180S}, a target-of-opportunity program has been proposed for follow-up observations with LSST that would require repeated observations in multiple filters of $\sim$100 deg$^2$ fields defined by GW localization regions  \cite{2019arXiv190205140S}. Such a program could be accommodated with greater flexibility and more observing time post-LSST, and would be able to use well established reference templates from the survey.  It would address several of the scientific priorities in Section~\ref{sec:science}, such as tests of cosmic acceleration at high redshift, and new tests of gravity.

A number of other scientific cases for continued operation of the observatory relate to follow-up observations of discoveries with LSST.  These include extending proper motion studies of the stellar populations of dwarf galaxies in the Milky Way halo, and  characterization of their dark matter content, which would advance the goal of understanding the fundamental nature of dark matter. Microlensing searches for astrophysical and primordial black holes would also benefit from the increased time baseline and observations of regions of high stellar density (i.e., the Galactic Bulge and/or the Magellanic Clouds).  

Another general science case for continuing LSST survey operations may be to obtain better overlap with other large-area deep optical surveys in support of joint cosmology analyses.  In particular, the planned 2000 deg$^2$ High Latitude Survey with the Nancy Grace Roman Observatory would be a prime example.  The specific region of the sky has not been defined yet, but the Roman observations will include near-infrared bands, and have angular resolution superior to LSST. The science case for joint processing, including refinement of photometric redshift estimates and improved measurement of weak lensing shear, has been thoroughly explored (e.g., \cite{2020arXiv200810663C}), and aligns with several of the priorities mentioned in Section~\ref{sec:science}.

We note that it is the post-LSST time period that aligns best with the anticipated access to Extremely Large Telescopes (ELTs), either the E-ELT in Europe or the US ELT Program. Follow-up observations of the faint objects discovered at Rubin, either during the LSST survey, or ``live'' during post-LSST operations, with ELTs is a key opportunity to capitalize on the investments made in Rubin. All of the above dark energy and dark matter science cases (and many more) would benefit from this synergy, whether through full exploitation of the LSST time delay lens and Milky Way dwarf satellite samples (improving the detailed modeling of these objects), higher-fidelity measurements of GW source kilonova properties, or the pinning down of the faint tail of the photometric redshift training set (which is needed for Roman as well as Rubin).

The costs for continuing Rubin operations post-LSST survey would be well understood and (by definition) would not involve instrument development or commissioning. The operating cost of Rubin is dominated by the labor needed to operate and maintain the summit and data facilities, with gradual efficiency gains being offset by natural escalation linked to the cost of living, such that to zeroth order the planned flat \$70M annual operating budget would need to be maintained into the post-LSST era. (Note that Rubin Observatory operations are jointly funded by NSF and DOE, with the balance in support determined by the science program. During the LSST survey, the operations costs are shared approximately equally.) Some reduction in operating cost could be possible if annual reprocessing of the LSST dataset is discontinued in favor of some less exhaustive approach (with knock-on effects on the labor needed to evolve the pipelines and support the science community), but this choice will be coupled to the science-driven but not-yet-known survey strategy. 

With the variety of strong cases for continued operation, and the variety of observing strategies that would be optimum for different cases, and in anticipation that the specific post-LSST science cases will be influenced by the findings from LSST, the detailed case for continued operation of Rubin, in terms of observing strategies, fields, and overall time needed, should be considered by year 5 of the survey.

\subsection{New Filters}

The baseline Rubin Observatory filter complement consists of six filters, \textit{ugrizy}, any five of which are resident in the camera at any given time.  The filters can be switched between individual observations with a changeout time of two minutes.  The sixth filter can be installed to replace any of the other five during daylight hours.  The filters were produced using interference coatings on polished, curved glass substrates.  The selection of the bandpasses was chosen early in the construction effort to satisfy a number of science requirements imposed by both astrophysical and cosmological science objectives.

After the LSST is completed, a new or partially modified filter set could be incorporated.  Fabrication of new filters would cost  about \$2M per filter, although some cost savings could possibly be achieved if some or all of the existing filter blanks were reused.

Possibilities for new filters could include:

\begin{itemize}
	\item A complementary filter set to the original six might be fabricated to improve photometric redshift estimates of the catalogued galaxy sample.  This could involve filter bands of comparable width, but shifted by half the band to eventually define twelve, rather than six effective wavelength bands. A set of medium-band filters could also serve the same purpose.
    \item A set of narrow-band filters could be fabricated to enable emission line surveys for particular lines at redshift $z = 0$, or to select samples of galaxies at a set of discrete redshifts. Preliminary analysis suggests that interference filters with passbands as narrow as 5--10 nm are achievable.  
    \item A set of patterned filters, which would enable multiple bandpasses to be sampled simultaneously across the field.  Multicolor photometry of individual variable sources could then be obtained nearly contemporaneously by dithering the field between exposures.
\end{itemize}

All of these options would require further study to evaluate their technical feasibility and scientific rationale.  In order for a new filter set to be available at the end of the nominal 10-year LSST, fabrication would need to begin roughly 5--7 years into the survey.

\subsection{A New Instrument}

The last option we briefly discuss is the construction of an instrument that would transform the Rubin Observatory into a truly new facility. A wide variety of reports~\cite{2013arXiv1311.2496M,2015osgb.book,2016arXiv161001661N, 2017arXiv170101976E} have highlighted the broad science opportunities that a wide-field spectrograph would provide to follow up the rich LSST imaging dataset; the value of such a capability for probes of cosmology was presented in the recent Astro2020 Decadal Survey report \cite{2021pdaa.book.....N}. In Ref.~\cite{2019arXiv190504669S} the idea is discussed to realize these exciting opportunities with the Rubin Observatory itself by constructing a multi-fiber spectrograph. Such an instrument could possibly exploit the large etendue of Rubin. However, as outlined in Ref.~\cite{2019arXiv190504669S}, this approach would have many technological challenges to overcome. The report listed technical tasks that were identified to either resolve some of the challenges or at least reduce the risks associated with them. To fully evaluate the feasibility of building a new instrument for the Rubin Observatory, a detailed design and technology study is required in the near future.

\section{Summary}

The next decade undoubtedly will be an exciting but also challenging period for cosmology. The Stage IV dark energy experiments hold the promise to unravel some of the mysteries of the dark universe. The findings in the next few years will guide the  directions that the field will pursue in 2030 and beyond and the Rubin Observatory post-LSST can continue to play a major role in these new endeavors. In this White Paper we outlined scientific opportunities with the Rubin Observatory to further explore the physics of the dark universe post-LSST, when Rubin's faint object capabilities will be well-matched to the Extremely Large Telescopes that should then be online. We described three scenarios for how the Rubin Observatory could be used after LSST: (i) modified observing strategy (only), (ii) new filter complement, (iii) a new instrument. They will require different investments and more detailed studies. These studies should be undertaken a few years into the LSST survey so that they can be informed by the scientific findings.

\bibliographystyle{JHEP.bst}
\bibliography{main.bib}

\end{document}